\documentclass[aps,nofootinbib,pra,amsmath,amssymb,twocolumn]{revtex4-2}
\pdfoutput=1
\usepackage{amssymb}
\usepackage{color}
\usepackage{graphicx}
\usepackage{epsfig,amssymb,amsmath}
\usepackage{chngcntr}
\usepackage[normalem]{ulem}

\graphicspath{ {./images/} } 
\usepackage[caption=false]{subfig} 
\usepackage{float}
\usepackage{dsfont}

\usepackage{color}
\newcounter{bar}

\newcounter{num}
\def\num{\par\medskip\refstepcounter{num}\hangindent4pt{\bfseries\arabic{num}}.~}

\newenvironment{numera}
{\parindent0pt\par
}
{\setcounter{num}{0}\par\smallskip}

\def\>{\rangle}
\def\<{\langle}

\def\labell#1{\label{#1}}
\def\section#1{{\par\em #1:--- }}
\def\togli#1{}

\begin{document}

\title{Quantum stroboscopy for time measurements} \author{Seth
  Lloyd$^1$, Lorenzo Maccone$^2$, Lionel Martellini$^3$, and Simone
  Roncallo$^2$}\affiliation{\vbox{$1.$ Massachusetts Institute of
    Technology, 77 Massachusetts Avenue, Cambridge MA,
    USA}\\{\vbox{$2.$~Dipartimento di Fisica and INFN Sezione di Pavia, University of
      Pavia, Via Bassi 6, I-27100 Pavia, Italy}\\\vbox{$3.$ EDHEC Quantum Institute, Nice, France}}} \pacs{}
\begin{abstract}
  Mielnik's cannonball argument uses the Zeno effect to argue that {\em projective} measurements for time of arrival are impossible. If one repeatedly measures the position of a particle (or a cannonball!) that has yet to arrive at a detector, the Zeno effect will repeatedly collapse its wavefunction away from it: the particle never arrives. Here we introduce quantum stroboscopic measurements where we accumulate statistics of projective position measurements, {\em performed on different copies of the system} at different times, to obtain a time-of-arrival distribution. We show that, under appropriate limits, this gives the same statistics as time measurements of conventional ``always on'' particle detectors, that bypass Mielnik's argument using non-projective, weak continuous measurements. In addition to time of arrival, quantum stroboscopy can describe distributions of general time measurements. It can also be adapted to obtain the conditional probability distribution of arrival times, given that the particle was not previously detected at the detector.
\end{abstract}
\maketitle

If one performs a projective measurement of an observable repeatedly
at a high rate, the first measurement will collapse the system to an
eigenstate and the subsequent ones will freeze the system to that
eigenstate, preventing further evolution: the Zeno
effect~\cite{zeno1,zeno2}. This seems to prevent the possibility of
describing time-of-arrival measurements, i.e.~the time at which a
particle arrives at the position of a detector, in the case 
in which it had a negligible probability of being at the detector 
position initially. Indeed, such a detector must measure the
position repeatedly with a high repetition rate $\tau$. But these
measurements will collapse the position of the particle away from the
detector, and the particle will never arrive there. Mielnik
\cite{mielnik} points out that, by the same argument, one can even
stop cannonballs!  Barchielli et
al.~\cite{barchiellinuovocim,barchiellilanz} showed that typical
detectors {\em are} able to detect particles, because they do not
perform exact projective measurements: they measure a position
``fuzzily'' \cite{peres}, with a variance $\sigma$ that must scale at
least as $1/\tau$ to avoid the Zeno effect. This scaling persists also
in the continuous limit $\tau\to0$, namely the product $\sigma\tau$
must be a constant $\kappa$, inversely proportional to the coupling
strength $1/2\kappa$ between the apparatus and the system
\cite{barchiellinuovocim}. In contrast, strong projective precise
measurements ($\sigma=0$) have infinite coupling to the apparatus
$\kappa\to\infty$. Many different models of continuous time
measurements (``always on'' detectors) of this type have been studied,
e.g.~\cite{peres,cavesmilburn,genoni,barchiellipaganoni,mandelmeltzer,mandelwolf}. These
are ``weak'' measurements, due to the finite coupling $\kappa$ or to
the fuzziness $\sigma$. They have been used to obtain time
distributions, e.g.~the ``waiting times'' \cite{vyas,vyas1}, or
\cite{molmer,steve}. However, the coupling to the apparatus alters the
dynamics of the system, yielding a master equation
\cite{qjump1,qjump2} with a dissipative/dispersive term of order
$1/\kappa$ \cite{barchiellinuovocim,barchiellilanz}, so that time
measurements outcomes are typically distorted in a way that cannot be
easily fixed.

\begin{figure}[htb]
\begin{center}
\epsfxsize=.75\hsize\leavevmode\epsffile{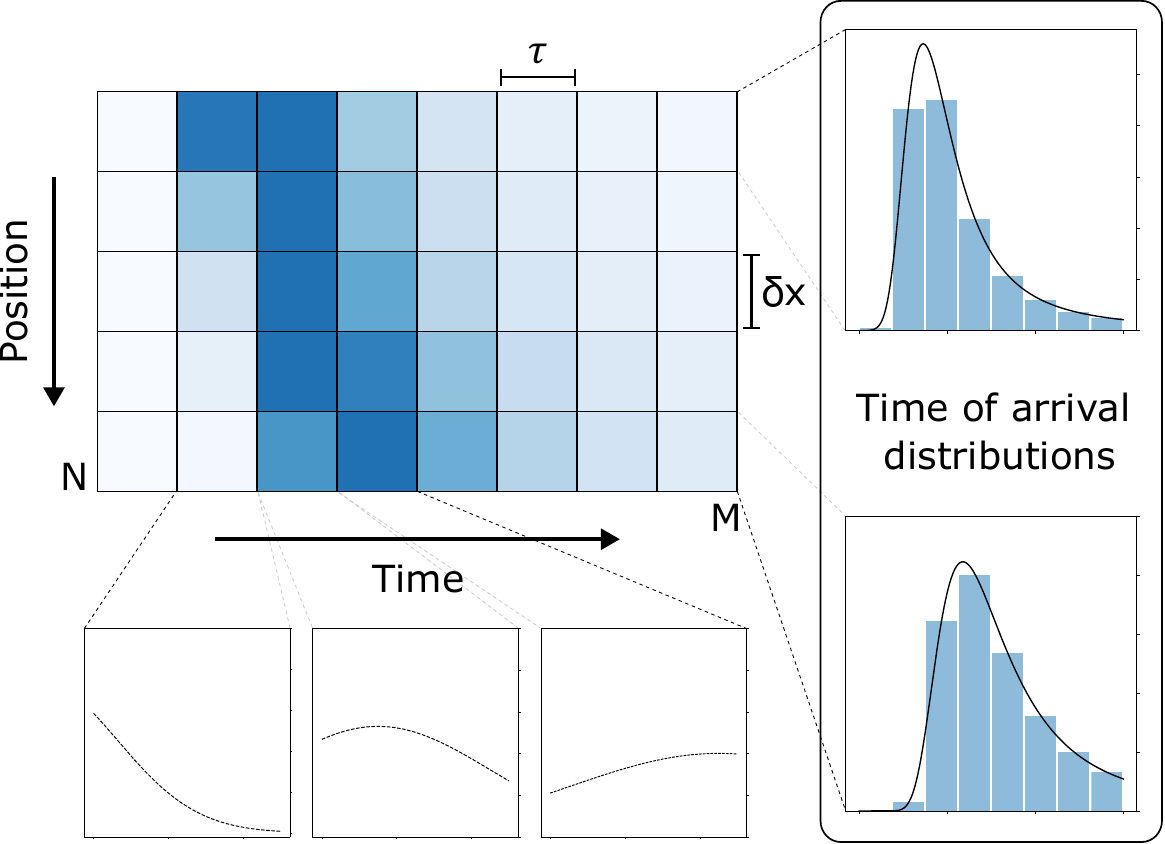}
\end{center}
\vspace{-.5cm}
\caption{Quantum stroboscopy for position. Divide the total
  measurement duration $T$ into $M$ intervals $\tau=T/M$. Then perform
  $L$ position measurements at times ${t}_m=m\tau+{t}_0$ (with
  $m=0,1,\cdots,M-1$) where $L\gg N$ is larger than the number $N$ of
  histogram bins at each time. Populate the $m$th histogram with the
  outcomes (vertical lines): add an event to the $n$th row, $m$th
  column for a position result $n\delta x$ (with $n=0,\cdots,N-1$) at
  time ${t}_m$ ($\delta x$ is the spatial resolution).  Normalize the
  {\em rows}, to obtain the time-of-arrival probabilities at positions
  $n\delta x$ as $p(t_m|n)= \ell_{nm}/\sum_m \ell_{nm}$, where
  $\ell_{nm}$ is the number of shots with outcome $n$ at time ${t}_m$.
  By construction, this is normalized over time for all values of
  $n$. [It is undefined if there are rows $n$ with no events: the
  particle never ``arrives'' at location $n\delta x$]. A similar
  procedure can produce time probabilities of arbitrary measurements.
  Here we simulated a Gaussian packet propagating downward.  The
  colormap and the lower graphs represent the probability $p(n|t_m)$
  for each outcome at $t_m$. The right graphs show the time
  probability $p({t}_m|n)$ (histograms) and their compatibility
  (proved in the main body) with the quantum clock (continuous line).} \labell{f:schema}\end{figure}

In this paper we introduce quantum stroboscopy: a procedure that
overcomes both the Zeno effect and the distortions of continuous
measurements, by leaving the system untouched up to a time ${t}_m$
when a strong (projective) measurement is performed. By repeating on
different identically prepared copies of the system for $M$ different
times ${t}_m$, and appropriately renormalizing the outcome statistics
(Fig.~\ref{f:schema}), one can obtain {\em time distributions} that
pertain to that measurement (time of arrival being one specific
case). Under the above hypothesis that there is a negligible
initial probability of finding the particle at the detector, the
leading tails of the obtained distribution will give the earliest
``time of arrival''. Our proposal differs from the stroboscopic
procedures of \cite{barkai,barkai1,nikolic,dahr}, where repeated
measurements are performed on the {\em same} system. Importantly, we
show that the obtained statistics matches the one of the continuous
measurements described above, in the limit where the effect of the
coupling to the apparatus can be removed by accumulating sufficient
statistics. Indeed, in continuous measurements, the apparatus
disturbance typically grows with time, so that the ratio between the
measurement variances with and without apparatus is linear in the
total duration of the experiment $T$, and it is proportional to the
apparatus coupling $1/\kappa$. One can counteract this variance by
repeating the procedure a number of times $M'\gtrsim {T}/\kappa$, so
that the variance ${T}/\kappa M'$ of the average becomes
negligible. This is the same scaling as quantum stroboscopy where
$M={T}/\tau$. Indeed, if one takes also into account the fuzziness
$\sigma$ due to the avoidance of the Zeno effect, one must have
$\kappa=\sigma\tau$, where the continuous measurements can be seen as
small-$\tau$ limit of a succession of $\tau$-separated measurements
with variance $\sigma$. To statistically remove the fuzziness
$\sigma$, one needs to repeat the procedure an additional
$M''\propto\sigma$ times, so that it is repeated
$M' M''\propto \sigma T/\kappa=T/\tau=M$ times, where we used
$\kappa=\sigma\tau$. Namely, quantum stroboscopy uses the same number
of resources (experiment repetitions) $\propto T$ as the conventional
``always on'' detectors. But stroboscopy has the advantage that it
does not introduce any difficult-to-remove distortions of the time
distributions due to the apparatus coupling.

We also show that the quantum stroboscopy time distributions match the ones
obtained from the recent ``quantum clock'' proposal for time
measurements \cite{arrival}. A modified version of quantum stroboscopy can also generate the ``quantum flow'' proposal for time
measurements~\cite{Beau24_2}. In the following we will focus mainly on
time-of-arrival measurements, but our results can be immediately
extended to other time measurements, e.g.~``the time at which a spin
is up'' \cite{arrival,lionel}.

Outline: we first introduce quantum stroboscopy; then we show that its
probabilities are the same as the quantum clock's ones
\cite{arrival}; using the models in \cite{cavesmilburn} and
\cite{genoni}, we then show that they also match the ones from
continuous measurements for time of arrival; and finally, we discuss more
general measurements.

\section{Quantum stroboscopy}
Quantum stroboscopy consists in leaving the quantum system unperturbed
up to a time $t_m$ when a projective, ideally instantaneous,
measurement is performed and then the system is discarded, as also shown in \cite{Beau24,lionel}. Repeating
the whole procedure for different times (whence the name
``stroboscopy'') and then normalizing the measurement results over
time, we obtain the conditional probability that the time is $t$ given
that the system has a specific value of the measured observable. 

In more detail, quantum stroboscopy consists in:%
\begin{numera}
  \num Prepare the system at time ${t}\!=\!0$ and let it evolve
  freely.  \num At time ${t}_m=m\tau+{t}_0$, with $m=0,1,\cdots,M-1$
  and ${t}_0$ the time of the first measurement, perform a projective
  (``strong'') measurement of the observable
  $A=\sum_{n=0}^{N-1}a_nP_n$, with $a_n$ and $P_n$ its eigenvalues and its
  projectors on eigenspaces, respectively. Then discard the system.  
  \num Repeat steps 1 and 2 on $LM$ copies of
  the system with $L\gg N$, to obtain $M$ accurate (strong) measurements of the
  observable $A$ at $M$ different times ${t}_m$ \cite{Beau24,beau33,lionel}.
  \num Distribute the outcomes on an $N\times M$ matrix
  (Fig.~\ref{f:schema}) whose element $\ell_{nm}$ is equal to the
  number of times the outcome $a_n$ was obtained at time ${t}_m$. The
  time probability distribution, the ``probability that time is
  ${t}_m$ given that $A$ has value $a_n$'', is obtained by normalizing
  the {\em rows} of the $\ell_{nm}$ matrix:
\begin{align}
  p({t}_m|a_n)=\ell_{nm}/\sum_{m=0}^{M-1}\ell_{nm}.
  \labell{prob}\;
\end{align}
If a row $n$ has no entries, then the probability is undefined: the
system is {\em never} measured at value $a_n$ during the whole time
interval $[{t}_0,{t}_0+{T}]$ when it was probed. The number
of points that fall into each histogram bin follows a multinomial
distribution and the statistical error of each bin can be estimated
as $\sim1/\sqrt{L}$, arising from the statistics of a binomial
distribution, where success/failure means falling inside/outside the
bin. Moreover, timing errors introduced by a non-instantaneous
detector can be described replacing projective measurements with positive 
operator-valued measures (POVMs, see App.~I) \cite{repository}.
\end{numera}
In the above procedure, the choice of the time interval
$[{t}_0,{t}_0+{T}]$ can be done using some prior information on the
system. For example, for periodic evolutions, $T$ should be larger than a
period; for time of arrival it must contain the predicted time the
particle will be at the detector: the time ${t}_0$ must be smaller
than the predicted arrival time of the first wavepacket tail and $T$
must be of the order of the wavepacket duration $\Delta t$; and so on.
If, instead, no prior information is present, one can quickly converge
to the interesting $T$ by starting with a large $T$ and adjusting it
rapidly by first running the above procedure with a small $L$ (the
binomial probability of finding an interesting outcome at some time
ensures that it is sufficient),
finding the interval where the outcomes are interesting (e.g.~the
particle has arrived). Then one can increase $L$ only for such
interval.

We now apply it to time-of-arrival. There are two
inequivalent reasonable definitions \cite{wernerreview,rovelli,leon,giannit,muga,mugareview,muga2,mielnik,werner,kijioski,holevo,dipankar,aharonov,leon1,galapon,galapon1,galapon2,siddhant,lionel,Beau24,beau33,
Beau24_2, LM}: (i)~the time at which a particle is detected at the
detector (quantum stroboscopy). In the case in which the probability
of being detected there is initially negligible, then the leading
tail of the obtained distribution gives a notion of the time of
``first arrival''; (ii)~``first passage time'', i.e.~the time at
which the particle is detected at the detector conditioned upon the
fact it had not been detected there before (conditional quantum
stroboscopy). In the first case, the observable $A$ is a projector
$P_x$ onto the detector position $x$, and $N=2$ corresponding
to ``the particle is ($n=0$) or is not ($n=1$) at the
detector''. In the second case, the above procedure must be
slightly modified to consider repeated measurements on the same
system, namely ``conditional quantum stroboscopy'' (see also
\cite{barkai}): \begin{numera}\num Prepare the system at time
$t=0$.\num At every time step $\tau$ perform a projective
measurement {\em on the same system} at the detector position
until the particle is detected, recording the time ${t}=m\tau$, or
until a cutoff time ${t}_{max}$ is reached.\num Repeat the
procedure on different copies of the system collecting the
statistics of detections $t$. The histogram of such $t$s will
provide the conditional probability of the time at which the
particle arrives at the detector, given that it was not detected
there previously. As discussed above, to avoid the Zeno effect,
the product $\sigma\tau$ must be constant (decreasing $\tau$ must
correspond to increasing the measurement fuzziness
$\sigma$).\end{numera}

\section{Quantum clock}
We now show that the procedure (i) gives the same distribution as the
``quantum clock'' proposal of \cite{arrival}, which uses the Bayes
rule to define the probability distribution for time $t$ given that
the outcome of the measurement of $A$ is $a_n$, as
\begin{eqnarray}
  p_{qc}({t}|a_n)=\<\psi({t})|P_n|\psi({t})\>/\int_{{t}_0}^{{t}_0+{T}}dt\:\<\psi({t})|P_n|\psi({t})\>
  \labell{probqc},
\end{eqnarray}
where $|\psi({t})\>$ is the system state at time $t$ and $T$ is the
total time that the procedure is run for, see Eq.~(9) of
\cite{arrival}. Consider the limit $L\to\infty$ of infinite
measurements at each time step, the Born rule implies that, for each
$m$, $\ell_{nm}/L\to\<\psi({t})|P_n|\psi({t})\>$.  Taking also
$M\to\infty$:
\begin{align}
  \frac \tau{L}\sum_{m=0}^{M-1}\ell_{nm}\to
  \int_{{t}_0}^{{t}_0+{T}}dt\:\<\psi({t})|P_n|\psi({t})\>
\labell{lim}\;,
\end{align}
where we used the fact that $\tau={T}/M\to dt$. Namely,
$\ell_{nm}/(\tau\sum_m\ell_{nm})\to p_{qc}({t}_m|a_n)$, so that
\begin{align}
p_{qc}({t}_m|a_n)dt  =\lim_{L,M\to\infty} \ell_{nm}/\sum_m\ell_{nm}
\labell{limit}\;,
\end{align}
where the $dt$ reminds us that $p_{qc}$ is a probability {\em
  distribution}, namely $p_{qc}({t}|a_n)dt$ is the (dimensionless)
probability that $t$ is in the interval $[{t},{t}+dt]$, whereas
$p_{qc}$ by itself has dimensions of ${t}^{-1}$, as is clear from
\eqref{probqc}. The quantum clock probability distributions were
derived under the assumption that one can quantize time
\cite{qtime}. However, the above equivalence with quantum stroboscopy
implies that this assumption is not really necessary.

A \textit{different} notion of arrival \textit{at} time $t$ uses the \textit{flow} of probability, i.e.~the marginal change over time of the number of particles that reached the detector position \cite{Beau24, Beau24_2, lionel, LM}. In this case, the arrival time distribution is $\pi _{x}\left( t\right) \propto | \, \partial_t \int_{-\infty }^{x}\rho _{t}\left( u\right) du \, |$, with $\rho_{t}\left( x\right)$ the distribution of the measured position at time $t$, given by the Born rule \cite{Beau24_2}. In quantum stroboscopy, this can be cast by replacing \eqref{prob} with $\widehat{\pi }_{x}(t_{m}) \propto 
|\,\sum_{n/x_n<x}\ell_{nm}-\ell_{n(m-1)}\,|$, normalized over all the possible times $t_m$. Similarly, $\pi_{x}$ can be obtained by taking $L$ and $M$ to infinity.

\section{Continuous measurements}
We now show that, for the unconditioned time of arrival, the quantum
stroboscopy distribution can be obtained from continuous position
measurements. To this aim we use the Caves-Milburn model
\cite{cavesmilburn} of continuous position measurement. While this is
a specific model, it is optimal: it satisfies the general bounds of
\cite{barchiellinuovocim,barchiellilanz}: $\kappa=\sigma\tau$ is
constant in the limit of $\tau\to0$.

The model uses a simple von Neumann measurement where the position $x$
of the particle is coupled to $M$ single particle memories with the
coupling time-dependent Hamiltonian
$H_{int}=\sum_{m=0}^{M-1}\delta({t}-m\tau)xp_m$, where $p_m$ is the
momentum of the $m$th memory. Each memory is initially prepared in a
Gaussian wavepacket with variance $\sigma$ (it induces an exact
position measurement for $\sigma\to0$). We consider nonzero $\sigma$:
a fuzzy position measurement. Taking the limit $\tau\to0$ of
continuous measurements, the effect of the measurement becomes a
non-unitary coupling to an environment (the memories) which can be
described by the master equation \cite{cavesmilburn}
\begin{align}
{d\rho_t}/{dt}=-\tfrac{i}\hbar[H_0,\rho_t]-\gamma(x\rho_t x-\tfrac12\{x^2,\rho_t\})
\labell{meq}\;,
\end{align}
with $\rho_t$ the state of the particle, $H_0$ its free Hamiltonian
(in the absence of coupling to the measurement apparatus), and
$\gamma=1/(2\kappa)$ with $\kappa=\lim_{\tau\to0}\sigma\tau$. Its
effect is a dissipative dynamics that diffuses the particle's
momentum: $\Delta^2p({t})=\Delta^2p_0({t})+\hbar^2t/2\kappa$, where
$\Delta^2p$ is the overall variance in the momentum at time $t$, and
$\Delta^2p_0$ is the variance due to the free evolution only
\cite{cavesmilburn}. In the free, mass $m_p$ particle case, \eqref{meq} gives
equations of motion $d\<x\>_t/dt=\<p\>_t/m_p$, $d\<p\>_{t}/dt=0$, and \cite{beau}
\begin{align}
\Delta^2x({t})=\Delta^2x_0+c_0t+\Delta^2p_0t^2/m_p^2+\hbar^2{t}^3/6\kappa m_p^2
\labell{spread}\;,
\end{align}
with $\Delta^2 x_0,\Delta^2 p_0$ the initial variances, and $c_0$ an
integration constant, App.~II. Both position and momentum
uncertainties grow with time (Heisenberg uncertainty notwithstanding)
because the evolution \eqref{meq} is non-unitary (dissipative), due to
the disturbance induced by the apparatus. It can be easily
counterbalanced by repeating the procedure $M$ times and averaging, so
that the variance on the average is reduced by a factor $M$ (central
limit theorem). In the absence of apparatus disturbance
($\kappa\to\infty$), Eq.~\eqref{spread} gives a quadratic position
spread. Recovering a quadratic spread in \eqref{spread}, i.e.~in the
presence of disturbance requires $M\propto t$, which recovers a
$\propto {t}^2$ variance in the average position. This just recovers
the scaling: there is a distortion of the recovered distribution
because it is not governed by the (expected) constant $\Delta^2p_0$
but by a different one. In contrast quantum stroboscopy does not
suffer from this distortion and will indeed capture the correct
$\Delta^2p_0 {t}^2/m_p^2$ scaling.

So, in the limit in which one can neglect these distortions the two
procedures both give a distribution whose variance grows
quadratically. Indeed, the continuous measurement obtains $M$ particle
trajectories $\<x_j\>_t$ (which have a nonzero probability when probed
with suitable test functions \cite{barchiellilanz}). One then must
invert them to find the times ${t}_{kj}$ when $\<x_j\>_{{t}_{kj}}$ is
equal to the detector position. Distributing them on a histogram, one
can obtain the probability that time is $t$ given that the particle is
measured at the detector position, the same as for quantum
stroboscopy. (From the average trajectory, one can also obtain other
(trajectory-based) notions of time of arrival,
e.g.~\cite{aharonov,siddhant}.)

Trajectories are not necessary: in the framework of continuous
measurements one can consider a measurement that continuously checks
whether  a particle has arrived at a certain location $x$
\cite{genoni,mandelwolf,barchiellibelavkin,wiseman}. For example, one can use
the following interaction between a localized mode $c_x$ of the
particle's field and a bosonic memory mode $b_t$ (a different memory
for each time $t$):
$H_{int}({t})=i\sqrt{\kappa}(c_xb^\dag_{t}-c^\dag_xb_{t})$
\cite{genoni}, a time dependent Hamiltonian which couples $c_x$ to
$b_t$ only in the interval $[{t},{t}+dt]$. It is convenient to
introduce modes $B_{t}\equiv b_{t}\sqrt{dt}$ which satisfy
$[B_{t},B^\dag_{t}]=1$. The effect of this coupling on the evolution
of the system is again a dissipative master equation \cite{genoni}
\begin{align}
  d\rho_{t}/dt=-\tfrac i\hbar[H_0,\rho_{t}]-\gamma(c_x\rho_{t} c^\dag_x-\tfrac12\{c^\dag_xc_x,\rho_{t}\})
  \labell{meq1}\;.
\end{align}
By monitoring the
number of photons $N_{t}$ in the memory modes $B_t$, one sees a stream
of clicks $N_t$: a photon $N_{t}=1$ in mode $B_t$ implies that the
particle has arrived at time $t$ (to first order in $dt$ one never
sees more than 1 photon in $B_t$) and the probability of seeing a
photon at $t$ is $\<N_{t}\>=\gamma dt\beta$ with
$\beta=$Tr$[\rho^{c}_{t} c^\dag_xc_x]$, where $\rho^c_t$ is the state
of the system {\em conditioned} by its past history of clicks, and
where $N_t$ describes a Poisson increment \cite{genoni}.  This implies that, if the particle
is with certainty at the detector position, namely $\beta=1$, for a
time $\Delta t$ (a particle with a rectangular time-of-arrival
profile), then the click statistic in that interval is Poissonian with
expectation value and variance $\gamma\Delta {t}=\Delta
{t}/2\kappa$. Typical (non-rectangular) situations will entail that the click
statistics will be a product of a Poissonian times the temporal
wavepacket profile. In both cases, to counteract the Poissonian
variance, one needs to repeat the experiment at least a number
$M\propto\Delta {t}/\kappa$ of times, where $\Delta t$ is the interval
when the distribution is substantially nonzero. (This $M$ guarantees,
for example, that the statistical fluctuations of the average time of
arrival will not be affected much by the Poissonian spread.)  The
outcome of this procedure will be a string of times ${t}_j$ at which
$N_{{t}_j}=1$. To obtain a probability, one must arrange them into a
histogram. In the limit in which the apparatus disturbance in
\eqref{meq1} can be neglected, it gives the probability that time is
$t$ given that the particle was detected at the detector position, the
same as quantum stroboscopy (which does not require such a limit by
construction: the particle is unperturbed until it is measured). If
${T}\simeq\Delta t$ (as discussed above), we find that the number of
times $M$ one must repeat the continuous measurements is again of
order ${T}$, the same as quantum stroboscopy.

The above arguments can be used to treat all continuous measurements
where the variance $\Delta^2 A_{\mbox{\tiny time step}}$ of the
considered observables $A$ at each time $t$ are independent and
constant at each time step ${t}\to {t}+dt$. Namely, measurements where
the error at different times is uncorrelated. In this case, it is clear that the error
over a finite interval $T$ grows linearly with $T$:
$\Delta^2A_{\mbox{\tiny tot}}={T}\Delta^2 A_{\mbox{\tiny time step}}$
and one can remove the apparatus disturbance by increasing the
statistics $M$ by choosing $M\propto T$, which, again, is the same
scaling as quantum stroboscopy.

\section{General time measurements}
We now consider measurements beyond time of arrival
\cite{arrival,lionel}, e.g.~the time at which a driven two level atom
is excited or the time at which a precessing spin is up. In the first
case, the continuous measurement scenario involves a highly nontrivial
interplay between the atom and the driving field that plays the role
of the apparatus (see Eq.~2.14 of \cite{bar94countproc}), which
prevents a straightforward deconvolution to recover the free-evolution
temporal characteristics, although a numerical optimization of the
apparatus specs recovers some of them \cite{peres}, e.g.~their
oscillatory behavior. Instead, a stroboscopic measurement can easily
recover the full free-evolution distribution of the ``probability that
time is ${t}_m$ given that the atom is excited'':
$p({t}_m|$excited$)=\cos^2(\Omega {t_m}/2)/\sum_{m=0}^{M-1}\cos^2(\Omega
{t_m}/2)$ (with $\Omega$ the Rabi frequency), which, for
${M\to\infty}$ gives the quantum clock distribution
$\cos^2(\Omega {t}/2)/\int_0^Tdt\cos^2(\Omega {t}/2)$
\cite{arrival}. It closely tracks the free evolution (Rabi flopping),
namely the ``probability that the atom is excited given that time is
$t$'': $p($excited$|{t})=\cos^2(\Omega {t}/2)$.

\begin{figure*}[th]
	\centering
	\subfloat{\includegraphics[width = 0.32\linewidth]{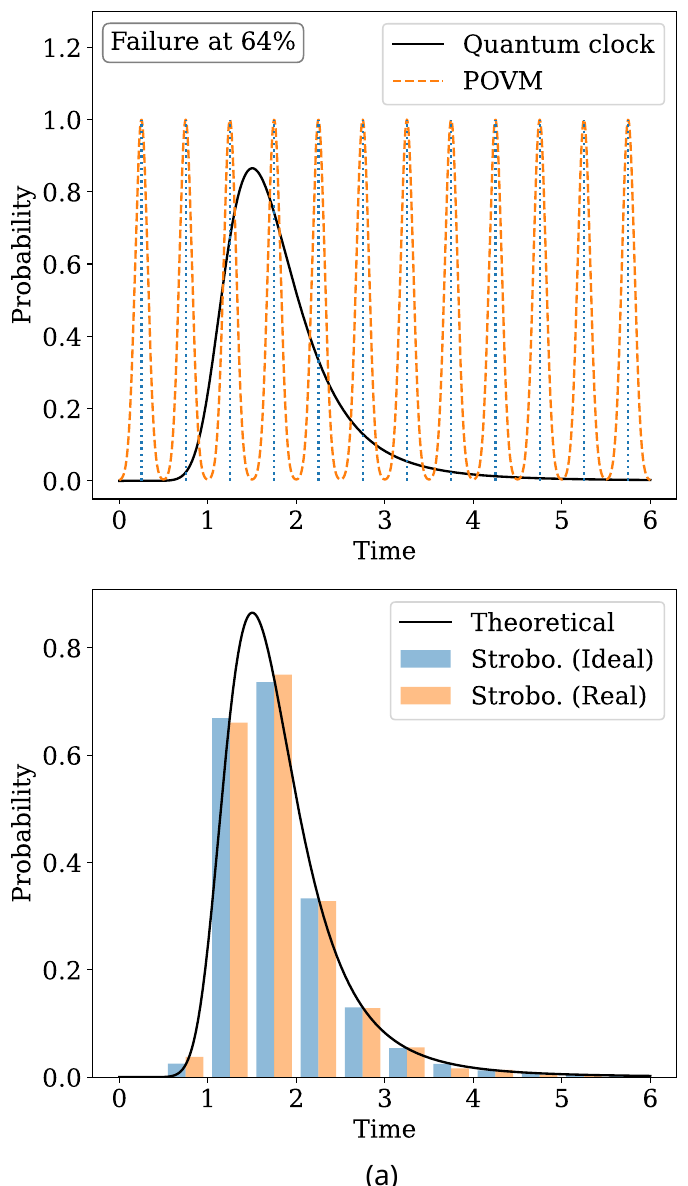}}%
   	\subfloat{\includegraphics[width = 0.32\linewidth]{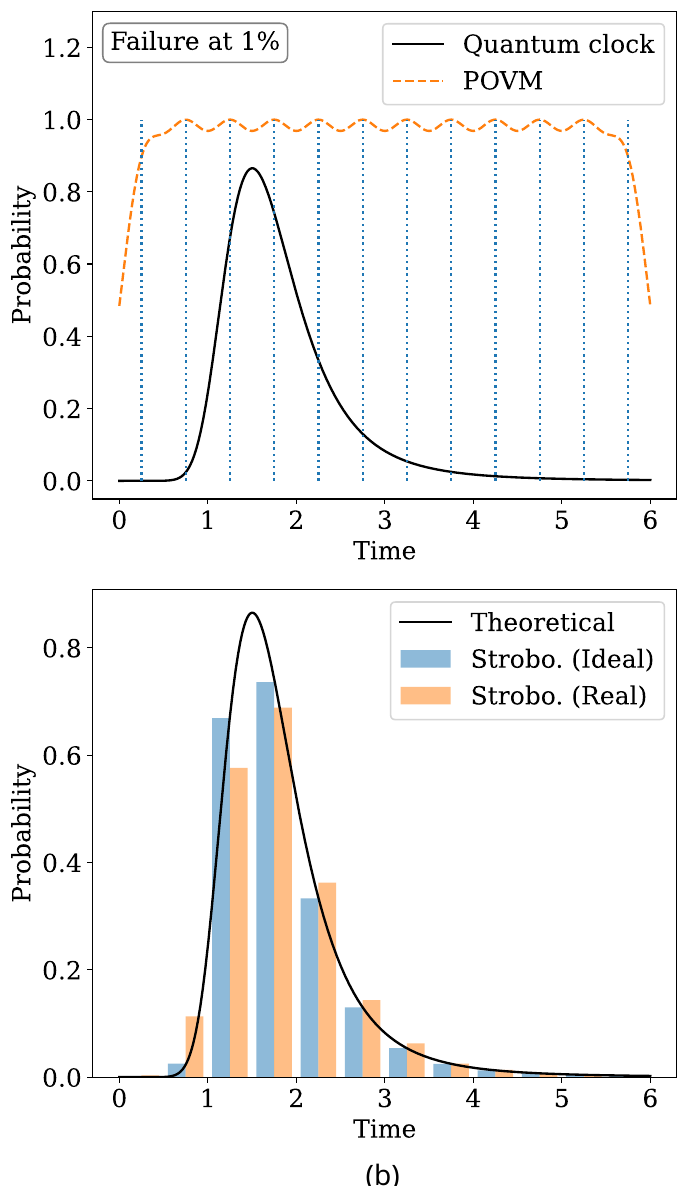}}%
   	\subfloat{\includegraphics[width = 0.32\linewidth]{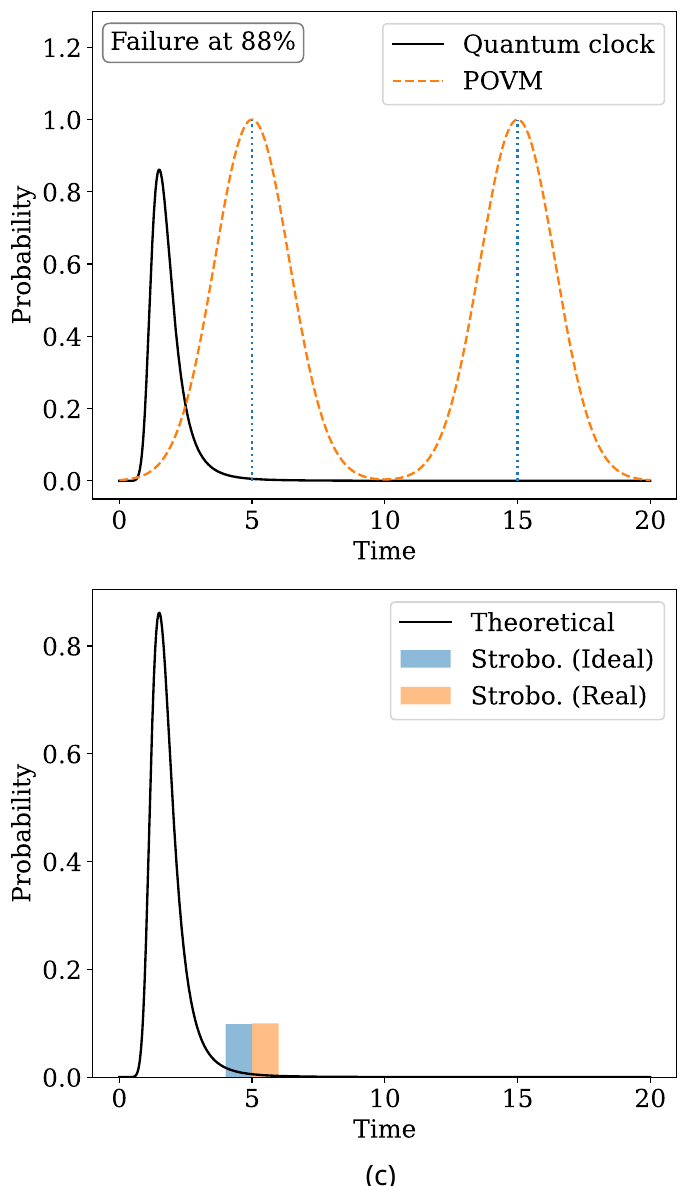}}%
    \caption{Monte Carlo simulation of non-instantaneous stroboscopic measurements. The upper panel shows the theoretical distribution (solid line), compared against the sequence of Gaussians that forms the POVM $\{\phi(m|t)\}$ (dashed line). Each vertical axis tracks a different stroboscopic bin. The lower panel compares the histograms of the ideal (projective) stroboscopic measurement against the non-instantaneous (POVM) one. (a) Stroboscopy for non-overlapping POVM. The real outcome closely resembles the projective one, with a $64\%$ probability of detection failure. (b) Stroboscopy with overlap. The real outcome significantly differs from the ideal one, with failure probability of $1\%$. (c) Worst-case scenario in which the POVM resolution is lower than the system time scale. Both the ideal and the non-instantaneous measurement collapse to a single bin.}
\end{figure*}

\section{Conclusions}
In conclusion, we have introduced quantum stroboscopy that produces
time distributions for measurements of arbitrary observables $A$.  It
returns the same time probability of the quantum clock proposal
\cite{arrival}, while a suitably modified version returns the time
probability of the ``quantum flow'' proposal~\cite{Beau24_2}. It also
returns the same distributions of the ``always on'' detectors, if the
intrinsic noise of such detectors can be bypassed using sufficient
statistics. In this case, the resource count of quantum stroboscopy
matches the one of the ``always on'' detectors: they both require a
number of repetitions $M$, linear in the total measurement time
$T$. Continuous measurements require $M$ repetitions to reduce the
apparatus error by averaging the outcomes, whereas quantum stroboscopy
requires $M$ repetitions because it is composed of $M$ projective
measurements at $M$ different times on $M$ copies of the system (one
measurement per copy). If, instead, the intrinsic noise of the
``always on'' detectors cannot be bypassed (as for Rabi flopping),
then these detectors become useless for time distributions, while
quantum stroboscopy still manages to correctly track the free
evolution. For time-of-arrival, quantum stroboscopy describes
the unconditioned distribution of finding the particle at the
detector. A slightly modified procedure, the ``conditioned quantum
stroboscopy'', returns the conditional distribution of finding the
particle at the detector given that it was not there previously
(first passage).

\section{Acknowledgements}
We acknowledge great feedback from Alberto Barchielli and Mathieu
Beau. S.R. acknowledges the PRIN MUR Project 2022RATBS4. L.M. acknowledges support from the National Research Centre for HPC, Big Data and Quantum Computing, PNRR MUR Project CN0000013-ICSC.

\section{Code availability}
The underlying code that generated the data for this study is openly available in GitHub \citep{repository}.

\vspace{1em}
\appendix\begin{centering}{\bf Appendix}\end{centering}

\section{I. Non-ideal detectors}
In this appendix, we analyze quantum stroboscopy for non ideal time measurements with finite resolution and timing errors. Timing errors typically consist of a jitter that introduces a random offset to each time measurement outcome, whereas a finite timing resolution implies that the measurement outcome has discrete values $\{{t}_m\}$ with integer $m$. We model such errors by replacing projective (strong instantaneous) measurements used in the main text with the POVM
\begin{gather}
	\big\{ \Pi_m = \int dt \phi(m|t) |t\>\<t| \ \text{with} \ m = 0,\cdots,M-1 \big\} \ , \\
	\Pi_{fail} = \mathds{1} - \sum_{m} \Pi_m \ ,
\end{gather}
where $\{\phi(m|t)\}$ is a sequence of equal-width Gaussians (whose widths represent the jitter) and whose averages identify the time bins $\{t_m\}$. Notice that $t$ and $m$ label the values of a continuous (time) and discrete (bin) random variable, respectively.

We consider different regimes where the POVM elements have significantly overlapping or disjoint supports. We analyze this scenario by performing a Monte Carlo simulation that consists of the following steps.
\begin{numera}
  \num Draw the ideal outcomes from the continuous stroboscopic distribution of Eq.~(1), e.g. via inverse random sampling or rejection sampling (whenever inverting the cumulative distribution is computationally challenging).
  \num For each sample, compute the cumulative distribution $F_\phi(m|t) = \sum_{\tilde{m}\leq m} \phi(\tilde{m}|t)$ under the normalization condition $\max_{t}\sum_{m}\phi(m|t) \leq 1$.
  \num Draw the real outcomes by applying inverse transform sampling on $F_\phi(m|t)$. In case of detection failure, i.e. when the particle is not collected by one of the bins, renormalize the distribution to successful events.
\end{numera}
Our simulations exhibit two different regimes. When the overlap decreases, the outcome distribution better approximates the ideal one with a non-negligible probability of failure. Conversely, a larger overlap reduces the probability of failure, but progressively dampens the distribution toward the uniform one. See Fig.~2.

\section{II. Derivation of Eq.~\eqref{spread}}
In this appendix, we analyze the diffusion of the particle's position
in the Caves-Milburn model, see also \cite{beau}. Consider the master equation describing the apparatus
disturbance, which can be equivalently written as
\begin{equation}
	d\rho_t/dt = -\tfrac{i}{\hbar}[H_0,\rho_t] - \tfrac{1}{4\kappa}[x,[x,\rho_t]] \ .
	\label{appeq:MasterEquation}
\end{equation}
Consider an observable $A$ with variance $\Delta^2 A = \langle A^2\rangle - \langle A \rangle^2$. In Schrödinger's picture, the observable expectation value evolves as $d \langle A \rangle / dt = \text{Tr}(A d\rho_t/dt)$, namely
\begin{equation}
	d\langle A \rangle_t / dt = - \tfrac{i}{\hbar}\langle[A,H_0]\rangle - \tfrac{1}{4\kappa}\langle[x,[x,A]]\rangle \ ,
	\label{appeq:Expectation}
\end{equation}
with variance $d\Delta^2 A_t/dt = d\langle A^2 \rangle_t /dt - 2\langle A \rangle_t d \langle A \rangle_t /dt$. From now on, we neglect the subscript $t$, which explicitly denotes time dependence, and we denote $I$ the identity operator. Since $[x,p^2] = 2i\hbar p$, the equations of motions (i.e.~the time evolution of the first-order moments) read $d\langle x \rangle /dt = \langle p \rangle / m_p$ and $d\langle p \rangle /dt = 0$. 

The second-order moment and the variance of $x$ and $p$ can be obtained as follows. From $[x,[x,p^2]] = -2\hbar^2I$ we get $d\langle p^2 \rangle /dt = \hbar^2/2\kappa$. By substitution and integration, the momentum variance reads
\begin{equation}
	\Delta^2 p = \Delta^2 p_0 + \hbar^2t/2\kappa \ ,
	\label{appeq:MomentumVariance} 
\end{equation} 
where $\Delta^2 p_0 $ accounts for the free evolution only. Moreover $[x^2,p^2] = 2i\hbar\{x,p\}$, yielding $d\langle x^2 \rangle /dt = \langle \{x,p\}\rangle /m_p$. Taking the second-order derivative of $\Delta^2 x$ gives
\begin{equation}
	d^2 \Delta^2 x /dt^2 = \tfrac{1}{m_p} d\langle \{x,p\}\rangle/dt - \tfrac{2}{m_p^2} \langle p \rangle^2  \ ,
	\label{appeq:PositionVariance}  
\end{equation}
where the second term is obtained by imposing the equations of motions. From $[\{x,p\},p^2] = 4i\hbar p^2$, we get $d\langle \{x,p\}\rangle/dt = 2 \langle p^2 \rangle / m_p$. From this, we obtain a direct relation between $\Delta^2 p$ (given by \eqref{appeq:MomentumVariance}) and the second-order derivative of $\Delta^2 x$, which integrated twice yields
\begin{equation}
	\Delta^2 x = \Delta^2 x_0 + c_0 t + \Delta^2 p_0 t^2/m_p^2 + \hbar^2t^3/6m_p^2\kappa \ ,
\end{equation}
with $\Delta^2 x_0, \ \Delta^2 p_0$ the initial variances, without disturbance, and $c_0$ an integration constant representing the initial position-momentum covariance. This is Eq.~\eqref{spread} of the main text.


\end{document}